\documentclass[aip,jcp,amsmath,amssymb,twocolumn,reprint]{revtex4}
\usepackage{hyperref}
\usepackage{graphicx,natbib,xfrac}
\usepackage{color,soul,dsfont,dcolumn}

\DeclareMathOperator{\diag}{diag}

\newcommand{\be}{\begin{equation}}
\newcommand{\ee}{\end{equation}}
\newcommand{\bea}{\begin{eqnarray}}
\newcommand{\eea}{\end{eqnarray}}
\newcommand{\bvec}[1]{\mathbf{#1}}
\newcommand{\transp}{^{\mathrm{T}}}

\newcommand{\Eq}[1]{Eq.~\eqref{#1}}
\newcommand{\Eqs}[1]{Eqs.~\eqref{#1}}
\newcommand{\Fig}{Fig.~\ref}

\newcommand{\Ft}{F_{\text{trial}}}
\newcommand{\br}{\bvec{r}}

\newcommand{\bK}{\bvec{K}}

\newcommand{\bM}{\bvec{M}}
   
\newcommand{\bq}{\bvec{q}}

\def\<{\left\langle}                 % expectation
\def\>{\right\rangle}                 % expectation

\begin{document}
\title{Water hexamer: Self-consistent phonons versus reversible scaling versus replica exchange molecular dynamics}
\author{Sandra E. Brown}
\author{Vladimir A. Mandelshtam}
\affiliation{Department of Chemistry, University of California, Irvine, CA 92697, USA}

\begin{abstract}

Classical free energies for the cage and prism isomers of water hexamer computed by the self-consistent phonons (SCP) method and reversible scaling (RS) method are presented for several flexible water potentials.  
Both methods have been augmented with a rotational correction for improved accuracy when working with clusters.  
Comparison of the SCP results with the RS results suggests a fairly broad temperature range over which the SCP approximation can be expected to give accurate results for systems of water clusters, and complements a previously reported assessment of SCP.  
Discrepancies between the SCP and RS results presented here, and recently published replica exchange molecular dynamics (REMD) results bring into question the convergence of the REMD and accompanying replica exchange path integral molecular dynamics results.  
In addition to the ever-present specter of unconverged results, several possible sources for the discrepancy are explored based on inherent characteristics of the methods used.

\end{abstract}

\maketitle

%------------------------------------------ INTRODUCTION ------------------------------------------%

\section{Introduction}
\label{sec:introduction}

The ability of small water clusters, and water hexamer in particular, to serve as model systems for elucidating the 
complicated structure and dynamics of bulk, condensed-phase water has made them the subject of intense and on-going interest.  
Water hexamer is the smallest water cluster whose minimum energy configurations exhibit three-dimensional 
structures similar to those found in bulk water, in contrast to the ring structures favored by smaller clusters 
\cite{GregoryClary1996, Xantheas2002,  Perez2012, SaykallyWales2012}.  
Its deceptively diminutive title of ``smallest drop of water" betrays the fact that water hexamer has proven to be 
a very challenging and even controversial system for experimentalists and theorists alike.  
On the experimental front, the problem has been attacked using a battery of spectroscopic techniques, 
but such spectroscopic data may be difficult to interpret reliably 
without the aid of complementary computational studies.  
Attempts to make structural assignments for experimental spectra with the help of simulations often resort to tactics 
such as uniformly shifting or scaling data, with little to no physical justification for doing so, 
e.g., uniformly shifting or scaling harmonic frequencies with the intention of correcting for anharmonicities 
in water so that a direct comparison with experiment can be made \cite{Tainter2012}.  
Inconsistencies and work-arounds such as these demonstrate that the current state of affairs leaves much to be desired.

In addressing the question of which isomer(s) of water hexamer are favored at a given temperature, an obvious approach would be the use of path integrals in conjunction with either Monte Carlo or molecular dynamics.  
This method is exact in the sense that the calculated values converge
to the true equilibrium values in the limit of large number
of beads used to discretize the path integral.   
It is then natural to question whether such simulations are
sufficiently long for the computed values, e.g., relative isomer
populations, to have converged to their true equilibrium values. 
Even so, it is not uncommon for assessments of convergence in
molecular dynamics or Monte Carlo simulations to be inadequate or
unreported altogether. 
For this reason the status of the reported results is often unclear,
while conducting an independent study in order to determine whether
previously reported results are converged are apt to be both
unrewarding and very difficult, if not impossible, to carry out, due
to an incomplete knowledge of the precise methodology used in the
original source.

Another possible choice, employed here, is the self-consistent phonons (SCP) method, introduced several
decades ago to incorporate anharmonic effects in an approximate
treatment of the nuclear dynamics of condensed phase systems \cite{Koehler1966, Gillis1968}.   
The last several years have witnessed a resurgence in the use of the
SCP method, particularly in the context of finite many-body systems.   
Notable results include recent calculations of the fundamental
frequencies of aromatic hydrocarbons \cite{Calvo2010} as well as
ground states of very large Lennard-Jones clusters \cite{Georgescu2011, Georgescu2012}.   
The SCP method maps a given many-body system localized in an energy minimum at thermal equilibrium 
to a reference harmonic system by optimizing the Helmholtz free energy in the framework of the Gibbs-Bogoliubov variational principle.  
Unlike path integral simulations, the SCP method does not suffer any challenges regarding numerical convergence.  
The trade-off to this advantage is that its inherently approximate nature makes the assessment of the SCP method's accuracy an issue of critical importance.

Here we aim to validate the (low-temperature) SCP results by comparison with reversible scaling \cite{KoningAntonelliYip1999, KoningCaiAntonelliYip2000} (RS) results.  
Originally developed as a non-equilibrium approach for determining free energy differences, here we utilize an equilibrium variation of RS which we have found to be more robust, and include a rotational correction for improved accuracy when simulating clusters.  
Like replica exchange molecular dynamics (REMD) and replica exchange path integral molecular dynamics (\mbox{RE-PIMD}), RS is an exact-in-principle method.  
A very favorable assessment of SCP has already been provided by Ref.~\onlinecite{Georgescu2013} 
for the very case of water hexamer, albeit for calculation of the ground state energies only (i.e., for $T = 0$\,K).  
Note that from both a numerical and a physical standpoint, 
the SCP approximation for a quantum system is essentially equivalent 
to that for the corresponding classical system at finite temperature, 
with the understanding that thermal and quantum fluctuations play similar roles.  
Therefore, demonstrating the accuracy of SCP for the classical case at finite temperatures would provide additional evidence in favor of the method's use for the more general quantum case.

In short, the goals of this work are two-fold:  first, to assess the accuracy and reliability of the SCP results by comparison with RS, and second, to explore the possible sources of disagreement between the RS and SCP results, and recently published REMD results \cite{Babin2013}.  
The methodological discussions and comparison of results carried out here in the classical regime are expected to help shed light on the more complicated situation occurring in the the quantum regime, in particular, the conflicting accounts of quantum and isotope effects in water hexamer, which is being investigated in detail elsewhere \footnote{S. E. Brown and V. A. Mandelshtam, Submitted to The Journal of Chemical Physics (2014)}.

%-------------------------------------------- METHODS ---------------------------------------------%

\section{Methods}
\label{sec:methods}

\subsection{Self-Consistent Phonons}

Given a many-body system localized in a minimum at thermal equilibrium at temperature $T$, described by the Hamiltonian
\begin{equation}
\hat H = -\frac{\hbar^2}{2}\nabla\transp\bM ^{-1}\nabla + V(\br ),
\label{Hsys}
\end{equation}
where $V(\br )$ defines the potential energy surface (PES), $\bM =\diag\{m_i\}$ is the mass matrix, and $\br$, 
the coordinate vector, 
the SCP approximation returns an effective temperature-dependent, harmonic Hamiltonian for the system
\be \label{Href}
\hat H_{h}(T) = -\frac{\hbar^2}{2}\nabla\transp\bM ^{-1}\nabla +
\frac{1}{2}(\br -\bq)\transp\bK\,(\br -\bq)+V_0\
, 
\ee
where $V_0$ is the minimum of the effective harmonic potential, $\bq$ is its center, and  $\bK$ its Hessian.  
This is achieved by minimizing the Helmholtz free energy of the harmonic system using the Gibbs-Bogoliubov inequality
\be
F\leq\Ft:=F_h + \langle V\rangle_h - \langle V_h \rangle_h \; ,
\ee
which yields a system of coupled, nonlinear differential equations
\begin{subequations}
 \label{scp:gen}
\begin{align}
&\< \nabla V \>_{h} =0\\
& \< \nabla\nabla\transp V \>_{h} \label{SCP:K}
 = \bK \, ,
\end{align}
\end{subequations}
from which the variational parameters $\bq$ and $\bK$ can be solved for iteratively, i.e., self-consistently.  
Here $\langle \cdot \rangle_h$ denotes an ensemble average with respect to the effective harmonic potential.  
By design, the SCP method is ideally suited for free energy calculations.  
We have recently demonstrated that the numerical bottleneck of the method, 
the accurate evaluation of Gaussian integrals corresponding to
the expectation value of the potential and its derivatives, can be
overcome by employing quasi-Monte Carlo integration in place of
standard Monte Carlo integration \cite{Brown2013}.  

Note that although we have formulated the SCP method for the general case of a quantum \mbox{$N$-body} system, the special case of a classical system ($\hbar=0$) does not lead to either conceptual or algorithmic simplification as the central numerical problem remains to be the calculation of Gaussian integrals that appear in the definition of the thermal averages (cf. \Eqs{scp:gen}).  Therefore, the present assessment of SCP for the classical system serves simultaneously as its assessment for the same quantum system.

The interested reader is referred to Refs.~\onlinecite{Georgescu2012} and \onlinecite{Brown2013} for a more detailed description of SCP.

\subsection{Reversible Scaling}

Consider an isomer of a classical ($\hbar = 0$) cluster corresponding to a relatively deep and stable
potential energy minimum, i.e., assume that it is separated from
the rest of configuration space by relatively large energy barriers. In the
absence of an external field the translations of the center of mass can be
separated, so we may consider the subspace $\mathbb{R}^{(3N-3)}$ that includes 
only the vibrational degrees of freedom  and the rotations of the whole cluster. Because the
potential energy $U(\mathbf{r})$ is invariant to the rotations,
the energy minimum is a three-dimensional manifold.  
Consider the basin of attraction corresponding to the chosen isomer,
which is a region, ${\cal A}\in\mathbb{R}^{(3N-3)}$, in the
rovibrational configuration subspace
surrounding this rotationally invariant manifold.
Since eventually we are interested in calculating the free energy difference
between two different isomers, it suffices to consider the
contribution given by the
configuration integral only:

\be\label{eq:FT}
\beta F(T)=-\log\left[\int_{\cal A} d \mathbf{r} \, e^{-\beta U(\mathbf{r})}\right] \; ,
\ee
with $\beta =1/k_{\rm B} T$.  
Assuming that at some reference temperature $T_0=1/k_{\rm B} \beta_0$ the free energy is known, we can write
\bea \label{eq:TI}
\beta F(T)&=& \beta_0 F(T_0)+\int_{\beta_0}^\beta d\beta' \, \frac
d {d\beta'} [\beta' F(T')]
\\
&=& \beta_0 F(T_0)+\int_{\beta_0}^\beta d\beta' \, \left\langle U\right\rangle_{\beta'}
\eea
or
\be\label{eq:qs}
F(T)=\frac {\beta_0}{\beta} F(T_0)+W_{\rm qs} (T) \;,
\ee

where the quasi-static work  is
\be\label{eq:ERS}
W_{\rm qs}(T)=\frac 1\beta \int_{\beta_0}^\beta d\beta' \, \left\langle
  U\right\rangle_{\beta'}
\ee
with the average defined by 
\be
 \left\langle U\right\rangle_{\beta'} := \frac
 {\int_{\cal A} d \mathbf{r}\, U(\mathbf{r}) e^{-\beta' U(\mathbf{r})} } {\int_{\cal A} d \mathbf{r} \,
    e^{-\beta' U(\mathbf{r})} } \;.
\ee
Equation \eqref{eq:TI} is a variant of the thermodynamic integration
relationship with $\beta$ being a generalized displacement variable, and the
integrand, the generalized force. The
RS method {\cite{KoningAntonelliYip1999,
KoningCaiAntonelliYip2000} computes the dynamical work done along
a non-equilibrium process as an estimator for the quasi-static work,
enabling instantaneous values of the driving force to be used rather than equilibrium
ensemble averages:   
\be\label{eq:Wdyn}
W_{\rm qs}(T) \approx W_{\rm dyn}(T)=\frac 1 \beta \int _0^t  dt' \frac {d\beta(t')}{dt'} U_0[\mathbf{r}(t')] \;,
\ee
where we consider a function $\beta=\beta(t)$ with $\beta(0)=\beta_0$,
which changes slowly with the simulation time $t$.  At each time step,
a new configuration is generated by Metropolis Monte Carlo, and the
generalized force $U_0[\mathbf{r}(t)]$ computed.

The key advantage of this method over equilibrium approaches (e.g, thermodynamic integration, adiabatic switching) is that the entire $F(T)$ curve can be obtained in a single simulation.  
In the limit of an infinitely slow process \Eq{eq:Wdyn} becomes exact.  
However, for any finite-time realization of this approach $W_{\rm dyn}(T)$ will suffer from both statistical
and systematic errors.  
The statistical error can be reduced by considering the dynamical work averaged over
multiple finite simulations, which, by the second law of thermodynamics,
always gives an upper bound for the quasi-static work:
\be
W_{\rm dyn}(T) > W_{\rm qs}(T) \;.
\ee
Having implemented both versions of RS described above, 
we found its equilibrium version, \Eq{eq:ERS}, to be more robust.  
All reported results were obtained by estimating the quasi-static work using \Eq{eq:ERS}.

The prerequisite for RS is the existence of a reference state for which the free energy 
of the system is known.  Here we have taken our reference to be the physical system 
at a sufficiently low temperature $T_0$ such that the free energy $F(T_0)$
can be estimated accurately using a harmonic (or quasi-harmonic) approximation.  
The free energy as a function of temperature $F(T)$ is then computed by evaluating the quasi-static work 
by integrating over the temperature, which changes gradually from $T_0$ to $T$.

While the standard normal mode/harmonic approximation was originally used to obtain a low-temperature free energy reference, we found it to be problematic for the present case where the Hessian is not available analytically, and must instead be evaluated using finite differences.  
Evaluation of the Hessian by finite differences was found to result in difficult to control errors in the normal mode frequencies, to which the free energy differences are extremely sensitive.  
This is because large cancellations result in a free energy difference which is generally much smaller in magnitude than the free energy values from which it is determined.  
Moreover, low frequency modes account for the greatest contribution to the free energy differences for the case of water clusters.  
Inaccurate reference values arising from using the standard harmonic approximation with a finite-difference Hessian would manifest as a change in the slope of $F(T)$, according to \Eq{eq:qs}.
In order to circumvent the problem, here the reference free energy is estimated by the SCP method, 
which does not require an explicit knowledge of the Hessian \cite{Brown2013}, and can in principle achieve 
any desired accuracy.  
In the $T_0\to 0$ limit, the standard harmonic approximation and the SCP approximation coincide.

The above procedure can be used to estimate the free energy difference between cluster isomers in the 
Eckart subspace, i.e., the ($3N-6$)-dimensional subspace orthogonal to the rotational
and translational degrees of freedom (see, e.g., the discussion in Ref. \onlinecite{Brown2013}).  
In this case we can set the value of $F(T_0)$ defined by a standard normal mode expression:
\be
F(T_0)\approx E_{\rm min}+\frac 1{\beta_0}\left[ \sum_{k=1}^{3N-6} \log (\beta_0\hbar\omega_k)
+ \log \sigma \right] \;, \ee
where we have dropped all terms that cancel when the free energy
difference is taken.  Here $E_{\rm min}$ denotes the energy at the minimum,
$\omega_k$ the effective harmonic frequencies, and $\sigma$ the order of the isomer point group.

\subsection{Rotational Correction}

The rotational contribution to the free energy may be important for
relatively small clusters and when the inertia tensors of
the two isomers in question are very different. In this case we propose to use a
rigid asymmetric top correction \cite{McQuarrieBook2000}, which, with the omission 
of the translational terms, yields

\begin{align}
F(T_0)\approx &\,E_{\rm min} 
\\\nonumber &+
\frac 1{\beta_0}\left[ \sum_{k=1}^{3N-6} \log (\beta_0\hbar\omega_k)
+ \log \sigma - \frac 1 2 \log \frac {I_1 I_2 I_3}{\beta_0^3\hbar^6} \right] \;,
\end{align}
where $I_1$, $I_2$ and $I_3$ are the principal moments of inertia of
the isomer evaluated at its minimum configuration.
Note that the same rotational term can be used to approximately
include the rotational contribution in the SCP free energy.
Also note that for water hexamer the rotational contribution to the
free energy difference between cage and prism is only $\sim 0.1 k_{\rm
  B}T$, which is much smaller than the discrepancy between our
present results and the REMD results \cite{Babin2013}.

To summarize, for two isomers $A$ and $B$ the free energy
difference at temperature $T$ can be estimated using 
\begin{align}
F_A & (T) - F_B(T) = E_{\rm min}^{A}-E_{\rm min}^{B}
\\\nonumber &+
\frac 1\beta\left[
\sum_{k=1}^{3N-6} \log \frac{\omega^A_k}{\omega^B_k} + \frac 1 2 \log
\frac{I^B_1I^B_2I^B_3} {I^A_1I^A_2I^A_3} + \log \frac {\sigma_A}{
  \sigma_B}\right]
\\\nonumber &+\frac 1\beta\left[ \int_{\beta_0}^\beta d\beta' \, \left( \left\langle
  U\right\rangle_{\beta'}^{\rm A} -\left\langle
  U\right\rangle_{\beta'}^{\rm B} \right)\right] \;.
\end{align}
Note that for a harmonic system the free energy difference is linear in $T$, and that it is the final term in the above expression that accounts for anharmonic contributions.

\subsection{Free Energy Differences from Replica Exchange Molecular Dynamics}

The replica exchange (RE) method \cite{SwendsenWang1986, Geyer1991, HukushimaNemoto1996} can be combined with either Monte Carlo or molecular dynamics simulations in order to overcome the problem of ``broken ergodicity", i.e., the effective inability of a simulation to sample minima of the PES which are separated by large barriers.  
Multiple ``replicas" of the system are initialized and maintained over a ladder of fixed temperatures $ T_1< \ldots < T_N$.  
Monte Carlo or molecular dynamics is carried out for each replica simultaneously, 
and replicas at two adjacent inverse temperatures $\beta_i, \beta_{i+1}$ on the ladder are allowed to periodically swap coordinates with acceptance probability 
\[
P_{\text{swap}} = \min [1, \exp \left\{ (\beta_{i+1} - \beta_i) (U({\mathbf r}_{i+1}) - U({\mathbf r}_i)) \right\} ] ,
\]
thereby coupling all replicas over the ladder while ensuring that detailed balance is satisfied.
In this manner, the ergodicity of the high temperature replicas in an REMD simulation makes all of the molecular dynamics trajectories ergodic.

Given the equilibrium populations of two isomers, $P_A$ and $P_B$,  at
some temperature $T$, 
e.g., obtained from an REMD simulation, the free energy difference at
this temperature can be determined using
\be\label{eq:DeltaF}
\Delta F= F_A - F_B = -k_{\rm B} T \log \frac{P_A}{P_B} \,.
\ee

%------------------------------------- RESULTS AND DISCUSSION --------------------------------------%

\section{Results and Discussion}
\label{sec:resultsanddiscussion}

SCP calculations were carried out following the protocol of Ref.~\onlinecite{Brown2013} 
for the two isomers of water hexamer of most interest, the cage and prism isomers, 
using the \mbox{q-TIP4P/F} \cite{Habershon2009}, WHBB \cite{Wang2011}, and \mbox{HBB2-pol} \cite{Medders2013} potentials.  
Both WHBB and \mbox{HBB2-pol} have been constructed by parametrizing high-level  {\it ab~initio} data, 
while \mbox{q-TIP4P/F} is an empirical potential.  
For the RS calculations we used its equilibrium variation in which the quasi-static work was computed by \Eq{eq:ERS}.  
Although the non-equilibrium calculations were also performed and resulted in similar results, we found the implementation of the non-equilibrium approach more cumbersome due to the need to optimize the parameter dependence $\beta=\beta(t)$ to reduce the systematic error.
The SCP result is exact for classical systems in the low-temperature limit and becomes approximate at higher temperatures.  
Therefore, the choice for the initial temperature $T_0$ in the thermodynamic integration procedure (Eq.~\eqref{eq:TI}) is somewhat arbitrary, as long as the resulting temperature dependence $F = F(T)$ is insensitive to $T_0$. We found the acceptable choice for $T_0$ for all three %four 
potentials to be anywhere in the range [2\,K, 10\,K].  Here we report the results using $T_0 = 5$\,K.  
The integration in \Eq{eq:ERS} was performed over a temperature grid with step $\Delta T = 1$\,K using ``Tai's model" \cite{Tai1994}.
The RS simulation is valid as long as the random walk stays in the basin of attraction corresponding to a particular isomer.  
At sufficiently high temperature it does leave this region. Consequently, the RS results are truncated at the corresponding temperatures, namely, $T \sim 50$\,K for the \mbox{q-TIP4P} 
potential, and $T \sim 75$\,K for WHBB and \mbox{HBB2-pol}.

The classical SCP and RS results are shown in Fig.~\ref{fig:all}.  
The present comparison of the SCP results with the exact-in-principle RS method identifies the suitable temperature range of the SCP approximation for the case of water hexamer.
Interestingly, for the two {\it ab~initio}-based potentials, WHBB and \mbox{HBB2-pol}, the agreement is very good for quite a large temperature interval, while the breakdown of the SCP approximation occurs at noticeably lower temperatures for \mbox{q-TIP4P/F}.   
This suggests that the \mbox{q-TIP4P/F} 
potential has a much rougher landscape and is much less harmonic than the WHBB and \mbox{HBB2-pol} potentials.

In a recent paper, Babin and Paesani (B\&P) reported the results of their simulations of water hexamer 
using \mbox{RE-PIMD} \cite{Babin2013}.  
In addition to the three potentials used here, B\&P also consider the empirical \mbox{TTM3-F} potential \cite{Fanourgakis2008}.   
Both empirical potentials are much less computationally expensive than the {\it ab~initio} potentials; 
WHBB is significantly more expensive than \mbox{HBB2-pol}.  
B\&P computed the populations of several isomers of classical (H$_2$O)$_6$ and quantum
(H$_2$O)$_6$ and (D$_2$O)$_6$ for all four potentials.  
Their work is a follow-up of an earlier publication \cite{Wang2012}, 
where the results for the WHBB water hexamer were first reported.

The populations of both classical and quantum
isomers as a function of temperature computed by \mbox{RE-PIMD} are displayed in 
Fig.~4 of Ref. \onlinecite{Babin2013}.  The corresponding free energy differences 
(for quantum isomers only) computed using \Eq{eq:DeltaF} are shown in Fig.~5 of Ref. \onlinecite{Babin2013}. 
A simple visual inspection of these figures reveals results which
appear reasonable and consistent with one another: smooth temperature
dependencies for all of the computed populations are observed, and the
changes in free energy due to quantum effects (i.e., between classical
(H$_2$O)$_6$ and quantum (H$_2$O)$_6$) and isotope effects (i.e.,
between (H$_2$O)$_6$ and isotopically substituted (D$_2$O)$_6$) are
what one would expect. 
Still, one must be cautious in accepting that such appealing results
are, in fact, converged and accurate.   
A key problem lies in the frequent exchanges between replicas,
resulting in distributions of isomers at different temperatures which
are highly correlated, and, as such, often display smooth and
physically reasonable temperature dependencies, 
subject to the overall replica distribution over the relevant isomers being stationary during the  entire course of simulation.
Moreover, the fact that the results remain unchanged even after the
simulation time has been increased significantly is no guarantee that
the results have fully converged to their true values.   
We refer the reader to Fig.~4 of Ref.~\onlinecite{Sharapov2007}, which
illustrates a striking example of apparent convergence with the use of
the RE method.  The figure demonstrates that even an exceptionally
long RE simulation, in which the results (heat capacities) exhibit a
smooth temperature dependence, and change only slightly using variable
simulation lengths, could in fact be very far from truly converged.

A prerequisite for the quantum \mbox{RE-PIMD} simulations of B\&P to be correct is that the
classical REMD simulations be converged.    
Here we have extracted the REMD populations of the prism and cage isomers directly from 
Fig.~4 of Ref.~\onlinecite{Babin2013} by digitizing the curves of the graph.  
The free energy differences $\Delta F = F_{\rm cage} - F_{\rm prism}$
were then computed by \Eq{eq:DeltaF} and included in \Fig{fig:all}.  
The data for $\Delta F$ is presented only for those
points for which both populations were sufficiently large to give a reliable result.  
Note that the WHBB free energy difference was also provided in Fig.~3
of Ref.~\onlinecite{Wang2012} and is consistent with that of \Fig{fig:all}.  
The present RS (and SCP) results disagree with the REMD results of B\&P, 
which suggests to us that the latter may not be converged.

\begin{figure*}
	\begin{centering}
	\includegraphics[width=1.0\linewidth]{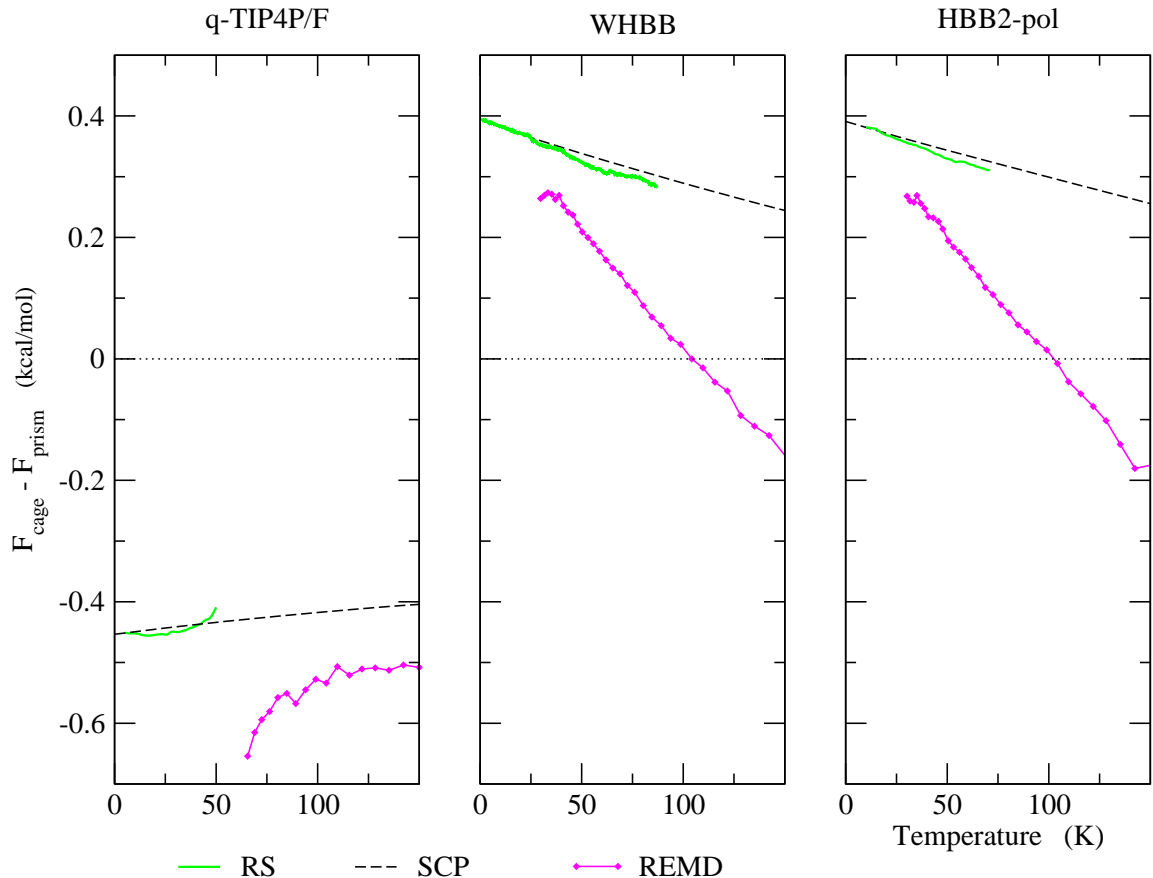}
	\end{centering}
	\caption{ \label{fig:all}
	Free energy differences for the cage and prism isomers of classical water hexamer computed by reversible scaling (RS) and self-consistent phonons (SCP).  RS results have been truncated at the temperature at which the method was found to break down (see discussion in the text).  Also included are the free energy differences derived from the populations of cage and prism isomers from the replica exchange molecular dynamics (REMD) simulations of Babin and Paesani \cite{Babin2013} using \Eq{eq:DeltaF}.
} 
\end{figure*}

While B\&P provide some discussion of their assessment of the convergence of the
\mbox{RE-PIMD} simulations, we believe that this assessment is not quite adequate.  
B\&P state that the convergence of their simulations was monitored by
ensuring that the ``round-trip times" over the RE temperature ladder were
significantly shorter than the overall simulation time, and by
analyzing much longer trajectories for the less computationally
expensive potentials.  However, the round-trip times are not
a true measure of ergodicity in the system. A correct measure of
ergodicity would be the characteristic time for the overall replica
distribution over the relevant isomers to fluctuate, which is
generally much longer than the round-trip time.

Perhaps the only reliable way to demonstrate that an RE simulation
is converged is to show that the results are invariant to the initial conditions.
Unfortunately, this is rarely done in practice, as it generally
requires one to carry out long equilibration calculations which may be
deemed too computationally demanding or unfeasible due to their length. 
This seems to be especially true of path integral simulations, 
including \mbox{RE-PIMD}, where a special effort to reduce equilibration
times is often made by employing a much less expensive classical simulation.  
In this scenario, the equilibrium results from the analogous classical
REMD simulation serve as the initial conditions for the quantum
simulation, under the assumption that this gives a distribution of
isomers over replicas which is not much different from the true
quantum equilibrium distribution of isomers.   
Although B\&P did not provide the reader with the complete details of
their numerical procedure, we believe that they likely initialized
each of their \mbox{RE-PIMD} simulations using the final replica distribution
from the corresponding classical REMD simulation.   
While this procedure of initializing the RE simulation is not ideal in
the sense that the close correlation between classical and quantum results 
may be an artifact of the choice of initial conditions, it provides a reasonable compromise 
for the case of a very expensive PIMD simulation.

Note that regardless of whether the isomer distribution
corresponds to the true equilibrium distribution or not, for an REMD
simulation which is stationary during the accumulation period, 
the free energy difference $\Delta F = F_{\rm cage} - F_{\rm prism}$
must approach the energy difference corresponding to the two energy
minima, $E_{\rm cage}-E_{\rm prism}$, in the $T\to 0$ limit.  
We emphasize that this condition only confirms that nothing 
non-stationary  happens during the accumulation period.  
In particular, the overall replica distribution over the relevant isomers must be stationary.  
This is certainly not the case for the \mbox{q-TIP4P/F} simulation, which,
independent of the comparison with our results, constitutes further evidence of its non-convergence.  
Note also that \mbox{q-TIP4P/F} is the least expensive potential of the four, for which the
longer 15~ns simulation time was used by B\&P (compared to 1~ns for the more expensive \mbox{HBB2-pol} and WHBB potentials).

Although RS can be used as an efficient alternative to the RE method, 
it is important to note that the optimal conditions for the RS and RE methods are opposite of one another, 
which can lead to difficulties when trying to compare results obtained using the two methods.  
In order for an RE simulation to be efficient (ergodic), the highest
replica temperature must be high enough for the random walk to switch
frequently between the relevant basins of attraction; an energy
barrier which is too high leads to a rapid loss of ergodicity.   
In the RS method, the random walk eventually leaves the basin of
attraction of interest at high enough temperature, leading to the
breakdown of the method.  
Like RS, SCP is designed to explore configurations which are confined to a particular basin of attraction of the PES.  
Given that there may be multiple minima of the PES which correspond to the same isomer, e.g. prism, this difference complicates the issue of making a direct comparison between the two types of methods.  
(Note that isomers are classified on the basis of the location of the oxygen atoms, while the positions of the hydrogen atoms may vary.)  
However, this issue can only account for the discrepancy between the SCP and RS results, and the REMD results under particular circumstances.  
Suppose that there are $N_{\text{A}}$ equivalent minima corresponding to isomer A, and $N_{\text{B}}$ equivalent minima corresponding to isomer B.  (We assume that the minima corresponding to a given isomer are equivalent for the sake of simplicity.)  
Then the partition functions for the isomers A and B computed by SCP or RS should be multiplied by $N_{\text{A}}$ and $N_{\text{B}}$, respectively.  
The correction to the free energy difference $F_{\text{A}} - F_{\text{B}}$ is then 
\[
- k_B T \log \frac{N_{\text{A}}}{N_{\text{B}}} \; .
\]
Clearly, this quantity can only become significant if the ratio $N_{\text{A}}/N_{\text{B}}$ is significantly different from unity.
Note also that this issue does not prevent B\&P from attempting to extrapolate their \mbox{RE-PIMD} results using the standard harmonic approximation, which also considers only a single basin of attraction, shown in Fig.~5 of Ref.~\onlinecite{Babin2013}.  
The related issue of assigning sampled configurations to a particular type of isomer when carrying out an REMD simulation is another possible source of discrepancy.  
B\&P have implemented a geometric criterion based on root mean squared distances, but it is unclear if this approach yields truly unambiguous identity assignments.

While an REMD simulation should, in principle, be able to sample multiple minima corresponding to the same isomer, 
it is difficult to determine if this was actually achieved in the case of Ref.~\onlinecite{Babin2013}.  
If we compare the details of their REMD simulation to those of previously reported simulations carried out by Tainter and Skinner (T\&S) 
using their classical empirical E3B potential \cite{Tainter2012}, then it seems questionable that B\&P could have achieved 
an ergodic simulation with converged results.  
T\&S used a temperature ladder ranging from 40~K to 194~K and a total simulation time of 250~ns, 
while B\&P report using a temperature ladder from 30~K to only 150~K, and total simulation times of only 1~ns for the WHBB and \mbox{HBB2-pol} potentials, and 15~ns for the \mbox{q-TIP4P/F} and \mbox{TTM3-F} potentials.  
T\&S also report that their simulation results were independent of the initial structures, 
and that the frequency of replica exchange remained low enough for the results to be independent of this frequency.  
B\&P appear to have been concerned only with ``round-trip" times.

%------------------------------------------- CONCLUSIONS -------------------------------------------%

\section{Conclusions}
\label{sec:conclusions}

We have presented evidence in favor of the accuracy of the SCP method for studying water clusters at low temperature based on comparison with the exact-in-principle RS method.  
Agreement between the two methods is seen within the temperature range up to 75 K for the more accurate WHBB and HBB2-pol potentials; we reiterate that SCP is exact for classical systems in the low-temperature limit.  
This agreement between RS and classical SCP also supports the use of SCP in the more general quantum case, given the similarities between thermal and quantum fluctuations both physically and numerically.

Additionally, we have presented evidence for the REMD and \mbox{RE-PIMD} results of Ref.~\onlinecite{Babin2013} 
(as well as the \mbox{RE-PIMD} results of Ref.~\onlinecite{Wang2012}) to be poorly converged.  
However, we cannot completely rule out the possibility of other
aspects contributing to the disagreement between the REMD and RS
results, including differences in sampling multiple basins of attraction corresponding to the same isomer, and assigning isomer identities to sampled configurations.  
Should the \mbox{RE-PIMD} results be, in fact, unconverged, this could account for the disagreement between recent studies of isotope effects in water hexamer, where SCP predicts small quantum and isotopic shifts relative to the energy differences between isomers, while the \mbox{RE-PIMD} results of B\&P predict quantum and isotope shifts which are sufficiently large to change the energy ordering of the cage and prism isomers.

\section*{Acknowledgements}
This work was supported by the National Science Foundation (NSF) Grant No.  \mbox{CHE-1152845}.  
SEB was partially supported by NSF Grant No. \mbox{DMS-1101578}.  
Volodymyr Babin and Francesco Paesani are acknowledged for discussing with us their results on water hexamer and for sending us the source code for the \mbox{HBB2-pol} PES.

%\appendix 
%\section{The Appendix}

%\bibliography{Papers/Comment}

\end{document}